\documentclass[11pt]{article}
\usepackage{mathrsfs}
\usepackage{amssymb}
\usepackage{tikz}
\usepackage{chet}
\usepackage{mathtools}

\newcommand{\CO}{\mathcal{O}}

\mathtoolsset{showonlyrefs=true}

\usetikzlibrary{arrows,calc}

\newcommand{\diag}{
\draw (0,0)--(1.5,0)
        (5,0)--(3.5,0)
        (2.5,1)--(2.5,-1);
  \draw (1.5,0) arc [start angle=180,end angle=150,radius=1cm]
  coordinate (left loop point);
  \draw (2.5,1) arc [start angle=90,end angle=120,radius=1cm]
  coordinate (right loop point);
  \draw let \p1=(left loop point), \p2=(right loop point) in
  (\x1,\y1) to [controls=+(120:0.3) and +(150:0.3)] (\x2,\y2);
  \draw let \p1=(left loop point), \p2=(right loop point) in (\x1,\y1)
  to [controls=+(330:0.3) and +(300:0.3)] (\x2,\y2);
  \draw (2.5,1) arc [start angle=90,end angle=45,radius=1cm] coordinate
  (right line point);
  \draw (3.5,0) arc [start angle=0,end angle=45,radius=1cm];
  \draw (1.5,0) arc [start angle=180,end angle=225,radius=1cm]
  coordinate (left line point);
  \draw (3.5,0) arc [start angle=360,end angle=225,radius=1cm];
  \draw let \p3=(left line point) in (\x3,\y3)--+(45:0.93cm);
  \draw let \p4=(right line point) in (\x4,\y4)--+(225:0.93cm);}

\makeatletter
\newlength\bothvec@height
\newlength\bothvec@depth
\newlength\bothvec@width
\newcommand{\bothvec}[2][]{%
    \settoheight{\bothvec@height}{$#2$}
    \settodepth{\bothvec@depth}{$#2$}
    \settowidth{\bothvec@width}{$#2$}
  \def\bothvec@arg{#1}
  \raisebox{.5ex}{\raisebox{\bothvec@height}{\rlap{%
    \kern.05em
    \begin{tikzpicture}[>=to]
    \pgfsetroundcap
    \coordinate (Stx) at (.05em,0) ;
    \coordinate (Arx) at (\bothvec@width-.05em,0) ;
    \draw[<->,very thin] (Stx)--(Arx);
    \end{tikzpicture}
  }}}
  #2
}
\makeatother

\SetFootnoteHook{\hangindent=0.5em\noindent}
\DeclareNewFootnote[para]{EE}[roman]
\renewcommand{\email}[1]{\footnoteEE{\href{mailto:#1}{\texttt{#1}}}}

\date{September 2012}

\preprint{CERN-PH-TH/2012-258\\
SU-ITP-12/30\\
UCSD-PTH-12-13}

\title{On Limit Cycles in Supersymmetric Theories}

\author{Jean-Fran\c{c}ois Fortin,$^{\ast,\dag,}$\email{jean-francois.fortin@cern.ch}
Benjam\'\i{}n Grinstein,$^{\$,}$\email{bgrinstein@ucsd.edu}
Christopher W.\ Murphy$^{\$,}$\email{cmurphy@physics.ucsd.edu} \\
and Andreas Stergiou$^{\$,}$\email{stergiou@physics.ucsd.edu}}

\affiliation{$^{\ast}$Theory Division, Department of Physics, CERN, CH-1211 Geneva 23, Switzerland\\
$^\dag$Stanford Institute for Theoretical Physics, Department of Physics, Stanford University, Stanford, CA 94305, USA\\
$^{\$}$Department of Physics, University of California-San Diego, La Jolla, CA 92093, USA}

\abstract{Contrary to popular belief conformality does not require zero
beta functions.  This follows from the work of Jack and Osborn, and
examples in non-supersymmetric theories were recently found by some of us.
In this note we show that such examples are absent in unitary $\mathcal
N=1$ supersymmetric four-dimensional field theories. More specifically, we
show to all orders in perturbation theory that the beta-function vector
field of such theories does not admit limit cycles.  A corollary of our
result is that unitary $\mathcal{N}=1$ supersymmetric four-dimensional
theories cannot be superscale-invariant without being superconformal.}

\begin{document}

\maketitle

\section{Introduction}
In recent papers by some of us two independent methods were used to claim
the existence of unitary four-dimensional quantum field theories that are
scale but not conformally invariant (SFTs)~\cite{Fortin:2011ks,
Fortin:2012ic, Fortin:2012cq}.  A natural interpretation of the
renormalization-group (RG) behavior of such theories is that they live on
RG limit cycles with a constant ``number of degrees of freedom.''
Nevertheless, the work of Jack and Osborn \cite{Jack:1990eb} (see also
\cite{Osborn:1991gm}), which we think is widely unappreciated in the
literature, has lead us to a new understanding of the conditions for
conformal invariance.\foot{We acknowledge helpful discussions on this point
with Markus Luty, Joseph Polchinski and Riccardo Rattazzi, as well as
informative correspondence with Hugh Osborn.}  More specifically, it has
become clear that a theory does not need to have zero beta functions in
order for it to be conformal, and that the claimed examples of
non-conformal scale-invariant field theories \cite{Fortin:2011ks,
Fortin:2012ic, Fortin:2012cq} are actually conformal.

We will not have much to say here about this new understanding---more
details will be given in a forthcoming publication \cite{Fortin:2012hn}.
Our aim in the present note is to show that unitary $\mathcal N=1$
supersymmetric theories in four dimensions cannot flow to a superconformal
phase with nonzero beta functions.  In other words, we will show that the
beta-function vector field of supersymmetric theories does not admit limit
cycles, in contrast to that of non-supersymmetric theories. (Let us remark
here that we use ``limit cycles'' loosely to mean recursive flows in the
beta-function vector field of a theory, that is, flows that may be cyclic
or ergodic.) A corollary of this result is that there are no unitary
$\mathcal{N}=1$ supersymmetric SFTs in four dimensions.

The subject of scale without conformal invariance in unitary $\mathcal N=1$
supersymmetric theories with an R-symmetry was investigated recently by
Antoniadis and Buican \cite{Antoniadis:2011gn}. Their treatment relies on
carefully analyzing constraints in the operator content of such theories,
and relies on various well motivated assumptions.  A criterion is then
given for a unitary supersymmetric theory to contain a superscale-invariant
phase: it has to contain at least two real nonconserved dimension-two
scalar singlet operators \cite{Antoniadis:2011gn}.  The most constraining
assumption in the analysis of \cite{Antoniadis:2011gn} is perhaps that an
R-symmetry is required along the RG flow.

The operator content of possible supersymmetric SFTs was also studied by
Nakayama \cite{Nakayama:2012nd}, without the requirement of an R-symmetry.
The so-called virial multiplet was constructed and its implications for
scale without conformal invariance in supersymmetric theories were
explored. In concrete examples difficulties were found in constructing a
nontrivial virial multiplet in perturbation theory. However, relaxing the
constraint of unitarity produced non-conformal scale-invariant field
theories in a simple Wess--Zumino model.

With the recent work mentioned in the last two paragraphs in mind,\foot{For
other studies of superscale and superconformal invariance see
\cite{Zheng:2011bp, Nakayama:2011tk}.} it seems unlikely that
supersymmetric theories can host a superscale-invariant phase that is not
superconformal.  Still, we think it is interesting to ponder the existence
of supersymmetric limit cycles.  Examples of limit cycles in
non-supersymmetric theories are more generic than previously thought: in
addition to a four-dimensional example, limit cycles in $4-\epsilon$
dimensions have also been found~\cite{Fortin:2011ks, Fortin:2011sz,
Fortin:2012ic}.  Thus, it is worthwhile to analyze the constraints
supersymmetry imposes on such RG behavior.

The conclusion of our present note is that supersymmetry does not allow for
limit cycles, and thus it does not allow for SFTs. Our method of proof, as
will become clear below, is very different in spirit from that employed by
Antoniadis and Buican, and by Nakayama. More specifically, in order to
reach our conclusion we analyze supersymmetric theories with
superspace-dependent couplings, and show that a quantity corresponding to
the $S$ of \cite{Jack:1990eb} (see also \cite{Fortin:2012hn}) is
constrained to be zero by supersymmetry.  The quantity $S$ is related to
the frequency with which a theory traverses its putative limit cycle, and
thus the fact that $S=0$ in supersymmetry immediately shows that
supersymmetric limit cycles cannot occur.

\textbf{Note Added:} As this work was being finalized, Nakayama added an
appendix to~\cite{Nakayama:2012nd} where he also showed that $S$ must
vanish to all orders in perturbation theory in $\mathcal N=1$
supersymmetric field theories.

\section{Preliminaries}\label{sec:prelim}
In this section we give a brief review of material that is necessary for
our arguments.

We are interested in four-dimensional theories that are classically
scale-invariant. They are parametrized by coupling constants $g_i$.
Following Jack and Osborn we promote these to spacetime-dependent
couplings, $g_i(x)$.  This is useful in two ways.  Firstly, the couplings
now act as sources for composite operators appearing in the Lagrangian.
This allows us to define finite composite operators as functional
derivatives of the renormalized generating functional for Green functions,
$W$, with respect to the couplings.  A similar method is used frequently to
define the stress-energy tensor: the theory is lifted to curved space and
the stress-energy tensor is obtained as a functional derivative of $W$ with
respect to the metric.  Secondly, it allows us to obtain a local version of
the Callan--Symanzik equation, with terms involving derivatives of
couplings interpreted as anomalies and thus satisfying Wess--Zumino
consistency conditions~\cite{Wess:1971yu}.

In order to render this theory finite one must include all possible
dimension-four counterterms consistent with diffeomorphism
invariance. In addition, the counterterms may be further constrained
by formal symmetries of the theory in which both quantum fields and
couplings transform. Consider, for example, a theory of real scalars
with bare Lagrangian
\eqn{\mathscr{L}_0=\tfrac12
\eta^{\mu\nu}\partial_{\mu}\phi_{0a}\partial_{\nu}\phi_{0a}
- \tfrac{1}{4!}g^0_{abcd}\phi_{0a}\phi_{0b}\phi_{0c}\phi_{0d}.
}[prebarescalarlag]
This is written in terms of bare fields $\phi_0$.  In the potential term
the bare couplings $g^0_{abcd}$ are completely symmetric under exchange of
the indices $a, b, c$ and $d$. The kinetic part of the Lagrangian exhibits
a continuous symmetry under transformations of the fields
$\delta\phi_{0a}=-\omega_{ab}\phi_{0b}$, where $\omega$ is in the Lie
algebra of the flavor group $G_F=SO(n_S)$. The whole Lagrangian is
$G_F$-symmetric if we agree to transform, in addition,  the couplings as
\eqn{\delta g^0_{abcd}=-\omega_{ae}g^0_{ebcd}-\omega_{be}g^0_{aecd}
-\omega_{ce}g^0_{abed}-\omega_{de}g^0_{abce},}
or $\delta g^0_I =-(\omega g^0)_I$ for short, where, following Jack and
Osborn, we use the compact notation $I=(abcd)$. For spacetime-independent
couplings the theory is renormalized by including the usual wave-function,
$\phi_0=Z\phi$,  and coupling constant, $g^0_I=g_I+L_I(g)$,
renormalization. But in the presence of spacetime-dependent coupling
constants one must introduce new counterterms. Among them we are
particularly interested in the counterterm of the form
\eqn{\mathscr{L}_{\text{c.t.}}=(\partial^\mu g_I)
(N_I)_{ab}\phi_{0b}\partial_\mu\phi_{0a},}[prenewct]
with $(N_I)_{ab}=-(N_I)_{ba}$, that is, in the Lie algebra of $G_F$; see
\cite{Jack:1990eb} for a complete account of counterterms required in the
case of spacetime-dependent couplings in a curved background.

Finite operators corresponding to currents associated with generators of
$G_F$ are most readily introduced by introducing background gauge fields.
We promote the Lagrangian \eqref{prebarescalarlag} to
\eqn{\widetilde{\mathscr{L}}_0=\tfrac12
g^{\mu\nu}D_{0\mu}\phi_{0a}D_{0\nu}\phi_{0a} +\tfrac1{12}\phi_{0a}\phi_{0a}
R - \tfrac{1}{4!}g^0_{abcd}\phi_{0a}\phi_{0b}\phi_{0c}\phi_{0d}\,,
}[barescalarlag]
where the covariant derivative,
\eqn{D_{0\mu}\phi_0=(\partial_\mu +A_{0\mu})\phi_0,}
is introduced with an eye towards including the counterterm \prenewct
through the renormalization of $A_{0\mu}$,
\eqn{A_{0\mu}=A_{\mu}+N_I(D_\mu g)_I,\qquad D_\mu=\partial_\mu+A_\mu.
}[bareAdefined]
We have left implicit the Lie-algebra indices (so that $N_I^T=-N_I$ and
$A_\mu^T=-A_\mu$). Note that $N_I$ is a function of the renormalized
couplings that has an expansion in $\epsilon$-poles starting at order
$1/\epsilon$.  If the theory contains gauge fields and some of the scalars
are charged under the gauge group $G_g\subseteq G_F$, it is straightforward
to include an additional quantum gauge field in addition to the background
field $A_\mu$.

The generating functional $W$ is now a function of the background gauge
field in addition to the metric and couplings, and finite operators are
defined by functional differentiation:
\eqn{\langle T_{\mu\nu}(x)\rangle=\frac{2}{\sqrt{-g}}\,
\frac{\delta W}{\delta g^{\mu\nu}(x)},
\qquad
\langle [\CO_i(x)]\rangle=
\frac{1}{\sqrt{-g}}\,
\frac{\delta  W}{\delta g^i(x)}\qquad\text{and}\qquad
\langle[\phi_a \bothvec{D}_{\mu}\phi_b]\rangle=
\frac{1}{\sqrt{-g}}\,
\frac{\delta  W}{\delta A^\mu_{ab}(x)}\,.}[Tmunudefined]
With this formalism Jack and Osborn obtain the trace-anomaly equation
\cite[Eq.~(6.15)]{Jack:1990eb}
\eqn{T^\mu_{\phantom{\mu}\!\mu}
=\beta_I[\CO_I]+ \partial^\mu[(\partial_\mu\phi)^T S \phi]
-((1+\gamma)\phi)\cdot\frac{\delta}{\delta\phi}S_0,}[TmumuGeneral]
where $S_0=\int d^4x\sqrt{-g}\,\mathscr{L}_0$ and $\beta_I$, and $\gamma$
are, as usual, the beta function of the coupling $g_I$ and the anomalous
dimension of the field $\phi$, respectively. We have specialized their
result to the case of flat metric, spacetime-independent couplings, and
vanishing background vector field.  The last term, involving the functional
derivative of the quantum action, vanishes by the equations of motion. The
surprising aspect of this result is the often neglected term that involves
the total divergence of the current $[(\partial_\mu\phi)^T S \phi]$. It is
defined in terms of the $G_F$-Lie algebra element
\eqn{S\equiv -g_IN^1_I,}
where $N_I=\sum_{n=1}^\infty N^n_I/\epsilon^n$, so that $N_I^1$ is the
residue of the simple $\epsilon$-pole in $N_I$. Moreover, using the
equation of motion (or the generalized symmetry under $G_F$) Jack and
Osborn get~\cite[Eq.~(6.23)]{Jack:1990eb}
\eqn{T^\mu_{\phantom{\mu}\!\mu}
=(\beta_I-(Sg)_I)[\CO_I]
-((1+\gamma+S)\phi)\cdot\frac{\delta}{\delta\phi}S_0.}[TmumuAgain]
This shows that a theory is conformal provided $\beta_I-(Sg)_I=0$. The
account above is readily generalized to the case of real scalars
interacting with Weyl fermions in the presence of quantum gauge fields.

In \cite{Fortin:2012hn} we used Weyl consistency conditions
\cite{Jack:1990eb, Osborn:1991gm} and perturbation theory to show that $S$
has two important properties:
\begin{enumerate}
\item $S$ vanishes at fixed points. That is, if $\beta_I=0$ then $S=0$.
\item On cycles, defined by $\beta_I=(Qg)_I$ for $Q$ in the Lie
  algebra of $G_F$, one has $S=Q$.
\end{enumerate}
Perturbation theory is only needed to establish positivity of the natural
metric in the space of operators, $\chi^g_{IJ}$ in the notation of
\cite{Jack:1990eb}.  It follows that in a theory for which $S=0$
identically there is no possibility of limit cycles, and that conformal
invariance corresponds to fixed points. We will show below this is
precisely the case for supersymmetric theories.

\section{Finding Limit Cycles} \label{sec:sscale}
In this section we review how to determine whether the beta-function vector
field of a theory admits limit cycles~\cite{Fortin:2012ic, Fortin:2012cq,
Fortin:2012hn}, making the procedure manifestly supersymmetric whenever
possible.  However, we often use what is known in the non-supersymmetric
case to deduce what conditions have to be satisfied in the supersymmetric
case.

Consider a classically scale-invariant supersymmetric field theory in four
dimensions with $N_f$ chiral superfields of mass dimension one.  Classical
scale invariance implies that the theory is renormalizable.  The part of
the Lagrangian we are interested in is\footnote{Lower case Roman letters
are indices in flavor space for (anti-)chiral superfields.}
\begin{equation} \label{eq:lag}
  \mathscr{L} = \int \! d^4\theta \, \Phi^{\dagger}_a
  \Phi^{\phantom{\dagger}}_a + \left(  \int
\! d^2\theta \, \frac{1}{3!} y_{abc} \Phi_a \Phi_b \Phi_c +
\text{h.c.}\right).
\end{equation}
There may be vector superfields in addition to the chiral superfields in
\eqref{eq:lag}, interacting in through a term $\Phi^\dagger_a
e^V\Phi^{\phantom{\dagger}}_a$ in the K\"{a}hler potential.  However, we do
not concern ourselves with vector superfields: their trivial flavor
structure renders them unable to play a role in determining whether limit
cycles exist.

The K\"{a}hler potential exhibits a continuous symmetry under
transformations of the fields $\delta\Phi_a=-\omega_{ab}\Phi_b$, where
$\omega$ is in the algebra of the ``flavor'' group $G_F = SU(N_f)$.  The
Yukawa couplings in the superpotential break $G_F$.  This flavor symmetry
can be extended to the whole Lagrangian by treating the coupling constants
as spurions, non-dynamical fields that are allowed to transform under
$G_F$.  More specifically, the coupling constant $y_{abc}$ is promoted to a
superspace-dependent chiral superfield of mass dimension zero,
\begin{equation}
Y_{abc}(z) = y_{abc}(z) + \sqrt{2} \theta y^{\psi}_{abc}(z) + \theta^2
y^F_{abc}(z),
\end{equation}
where $z^{\mu} = x^{\mu} + i \theta \sigma^{\mu} \bar{\theta}$.  The
$y^{\psi}$ and $y^F$ components of the spurion field are irrelevant and we
ignore them in what follows.  The Lagrangian~\eqref{eq:lag} is manifestly
$G_F$-symmetric if the Yukawa couplings transform as
\begin{equation}
\delta Y_{abc} = - \omega_{aa^{\prime}} Y_{a^{\prime}bc} -
\omega_{bb^{\prime}} Y_{ab^{\prime}c}  - \omega_{cc^{\prime}}
Y_{abc^{\prime}}.
\end{equation}

The theory also possesses a spurious $U(1)$ R-symmetry in addition to the
$G_F$ symmetry.  The fields and couplings transform under the R-symmetry
as
\begin{equation} \label{eq:rsym}
\Phi \rightarrow e^{i \alpha}\Phi,\quad \Phi^{\dagger} \rightarrow e^{- i
\alpha}\Phi^{\dagger},\quad Y \rightarrow e^{- i \alpha}Y,\quad
\overline{Y} \rightarrow e^{i \alpha}\overline{Y}.
\end{equation}
The R-symmetry is non-anomalous because the R-charge of the fermionic
component of $\Phi$ is zero.

We now look for a supersymmetric version of the new type of
counterterm that is required in the presence of superspace-dependent
couplings, as in \eqref{prenewct}.  In supersymmetric theories the
only candidate for this
counterterm has the form
\begin{equation} \label{eq:ct}
\mathscr{L}_{\text{c.t.}} = \int \! d^4\theta \, \Phi^{\dagger}_a F_{ab}
\Phi_b,
\end{equation}
where $F_{ab}$ is a function of the couplings.  If the theory is to be
unitary, $F_{ab}$ must be Hermitian, $F_{ab}(Y,\overline{Y}) =
F_{ba}(\overline{Y},Y) = F^*_{ba}(Y,\overline{Y})$.  One can readily
check that one of the components of \eqref{eq:ct} is of the form
\eqref{prenewct}, that is, the product of the current associated with $G_F$ and
the derivative of the couplings
\begin{equation} \label{eq:ctcomp}
\mathscr{L}_{\text{c.t.}} \supset \left((N_I)_{ab} \partial^{\mu}y_I -
(N_I)^*_{ba} \partial^{\mu}y^*_I \right)\left(\phi^{*}_a \partial_{\mu}
\phi_b - \partial_{\mu} \phi^{*}_a\, \phi_b\right),
\end{equation}
with $I$ again a shorthand for contracted flavor indices.  $N$ can be expressed
in terms of $F$ as
\begin{equation} \label{eq:N}
(N_I)_{ab} = \frac{\partial F_{ab}(y,y^*)}{\partial y_I}, \qquad (N_I)^*_{ba} =
\frac{\partial F_{ab}(y,y^*)}{\partial y^*_I}.
\end{equation}
Both $N$ and $F-1$ are functions of the renormalized couplings that have
$\epsilon$-pole expansions starting at order $1/\epsilon$.

\section{Absence of Limit Cycles in Supersymmetric Theories}
We are finally ready to prove at the quantum level that a unitary,
$\mathcal N=1$ supersymmetric field theory in four dimensions does not have
limit cycles.    Our strategy is to show that $S$ is exactly zero in
supersymmetric theories with the aforementioned qualifications. This we can
show without recourse to perturbation theory. However, we are mindful that
the proof in  \cite{Fortin:2012hn} that $S=Q$ on cycles and $S=0$ at fixed
points does rely on perturbation theory.

The expression for $S$ in our case is
\begin{align}
S_{ab} &\equiv -\tfrac{1}{2} (N^1_I)_{ab} y_I - \text{h.c.}, \label{eq:sdef}
\\ &= -\frac{1}{2} \left(y_I \frac{\partial F^1_{ab}(y,y^*)}{\partial y_I} - y^*_I
\frac{\partial F^1_{ab}(y,y^*)}{\partial y^*_I}\right), \label{eq:sab}
\end{align}
where $F^1$ is the residue of the simple $1/\epsilon$ pole in $F$.  The
Hermitian conjugate is subtracted in \eqref{eq:sdef}, as expected since
$S$ is anti-Hermitian.   The quantum action is invariant under the
R-symmetry introduced in Section \ref{sec:sscale}, see \eqref{eq:rsym}.
Therefore
\begin{equation} \label{eq:al}
F_{ab}(Y,\overline{Y}) = F_{ab}(e^{- i \alpha}Y,e^{ i \alpha}\overline{Y}),
\end{equation}
or, by taking $\alpha$ to be infinitesimal,
\begin{equation}
0 = Y_I \frac{\partial F_{ab}(Y,\overline{Y})}{\partial Y_I} - \overline{Y}_I
\frac{\partial F_{ab}(Y,\overline{Y})}{\partial \overline{Y}_I}.
\end{equation}
Comparing the scalar component of this equation with \eqref{eq:sab}
shows $S=0$.  The theory cannot exhibit renormalization group limit
cycles. Furthermore, unitarity and superscale invariance imply
superconformal invariance in unitary four dimensional $\mathcal N=1$
supersymmetric field theories.

\newsec{A Perturbative Proof and a Four-Loop Example}
If $S$ vanishes in supersymmetric theories non-perturbatively, the
implication must also be true to all orders in perturbation theory.  In
this section we illustrate the vanishing of $S$ in perturbation theory with
a four-loop example. Remarkably, four-loop calculations in the Wess--Zumino
model exist in the literature~\cite{Avdeev:1982jx}.  For a diagram
containing only chiral superfields, it is a simple combinatoric exercise to
convert the results of \cite{Avdeev:1982jx} to the model under
consideration in this work.

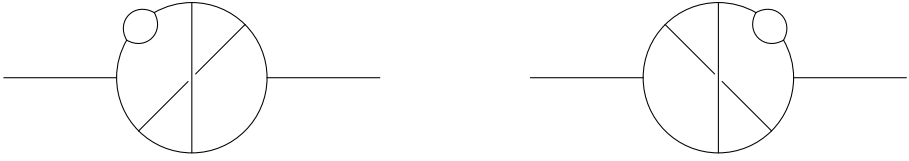
\begin{figure}[ht]
  \centering
  \begin{tikzpicture}[line cap=round,line join=smooth]
    \begin{scope}
      \diag
    \end{scope}
    \begin{scope}[xshift=12cm,yscale=-1,rotate=180]
      \diag
    \end{scope}
  \end{tikzpicture}
  \caption{Four-loop diagrams that contribute to $F$ that are
    asymmetric under exchange of the external legs.  The lines are
  superfield propagators.} \label{fig:4loop}
\end{figure}

In non-supersymmetric theories a scalar-propagator loop correction
contributes to $S$ if the corresponding diagram is not symmetric under $a
\leftrightarrow b$.  Such diagrams first arise at the three-loop level in
ordinary field theories~\cite{Fortin:2012hn}.  In $\mathcal N=1$
supersymmetric Wess--Zumino models asymmetric diagrams arise at four loops,
see e.g.\ Fig.~\ref{fig:4loop}.
The four-loop contribution of the diagrams of Fig.~\ref{fig:4loop} to $F^1$
is
\begin{multline}
(16\pi^2)^4 F^1_{ab}\supset \tfrac38(\zeta(3) -
\tfrac12\zeta(4))(y^{\phantom{*}}_{acd}
y^*_{dkm} y^{\phantom{*}}_{fk\ell} y^*_{bef} y^{\phantom{*}}_{ejm}
y^*_{ij\ell} y^{\phantom{*}}_{ghi} y^*_{cgh}\\ + y^{\phantom{*}}_{acd}
y^*_{dkm} y^{\phantom{*}}_{ik\ell} y^*_{ghi} y^{\phantom{*}}_{fgh}
y^*_{bef} y^{\phantom{*}}_{ejm} y^*_{cj\ell}),
\end{multline}
where $\zeta$ is the Riemann zeta function.  From this expression for $F^1$
we see that $S$ vanishes by \eqref{eq:sab}. There are at least two ways to
understand this diagrammatic result.

It is obvious from the form of \eqref{eq:sab} that $S$ counts the
difference in the number of $y$'s and $y^*$'s in $F$.  The
non-renormalization of the superpotential guarantees that any diagram
containing an unequal number of $y$'s and $y^*$'s vanishes.  Thus, the only
diagrams that contribute to $F$ contain an equal number of $y$'s and
$y^*$'s, and $S$ must vanish to all orders in perturbation theory. In
contrast with the non-supersymmetric case, not even diagrams asymmetric
under exchange of the external legs can contribute to $S$.

The second way in which our result can be understood is as follows.  In
non-supersymmetric theories momentum is allowed to flow into the diagram
that gives $N^1$ from an external leg and out of the diagram through a
coupling.  If the diagram is asymmetric, then interchanging the external
lines of the diagram results in a different routing of the external
momentum through the diagram, and thus to a different numerical coefficient
for the corresponding contribution to $N^1$. This leads to a nonzero
contribution to $S$ after antisymmetrization.  In the supersymmetric case,
however, the coefficient of all diagrams contained in the
$\theta$-expansion of an asymmetric diagram---like the one in
Fig.~\ref{fig:4loop}---comes from the zeroth-order in $\theta$ diagram,
which is calculated with no external momentum flowing into the diagram.
Thus, there is no possibility of a contribution to $S$.  This is true to
all orders in perturbation theory.

\ack{We thank Ken Intriligator and Hugh Osborn for useful discussions.  We
are especially grateful to Markus Luty, Joseph Polchinski and Riccardo
Rattazzi for constructive discussions.  BG, CM, and AS are supported in
part by the U.S.\ Department of Energy under contract No.\
DOE-FG03-97ER40546.  JFF is supported by the ERC grant BSMOXFORD No.\
228169.}

\newpage
\bibliography{Paper4dSUSY_ref}

\end{document}